\documentclass[11pt,a4paper]{article}

\usepackage[T1]{fontenc}
\usepackage{geometry}
\usepackage{amsmath,amssymb}
\usepackage{graphicx}
\usepackage{xcolor}
\usepackage{booktabs}
\usepackage{multirow}
\usepackage[version=3]{mhchem}
\usepackage{bm}
\usepackage{xurl}% long access URL in Data availability must break anywhere
\usepackage[super,numbers,sort&compress]{natbib}
\usepackage{hyperref}
\hypersetup{colorlinks=true, allcolors=blue}

\begin{document}

\begin{center}
  {\LARGE\bfseries Atomic-scale structure of static screening in noble-metal
   nanoparticles from clusters to the conductor limit\par}
  \vspace{1.4em}
  {\large Pulkit Joshi\textsuperscript{*}\quad Marek Sierka\textsuperscript{*}\par}
  \vspace{0.8em}
  Otto Schott Institute for Materials Research, Friedrich Schiller University,\\
  L{\"o}bdergraben 32, 07743 Jena, Germany\par
  \vspace{0.6em}
  {\small\textsuperscript{*}Corresponding authors:
   \texttt{pulkit.joshi@uni-jena.de}, \texttt{marek.sierka@uni-jena.de}\par}
\end{center}

\begin{abstract}
The static screening response of a metallic nanoparticle is encoded in its induced charge density. Integrated observables such as the dipole moment or the polarizability do not determine this density, because many different screening profiles yield the same integrals. We show that in noble-metal nanoparticles the screening charge is atomically structured, with facet, edge, and vertex sites responding differently from a smooth classical conductor. This structure survives in a surface layer of about one atomic width even when the integrated response has reached the conductor limit, and it is absent from continuum descriptions. To resolve it across size, morphology, and composition, we calibrate an atomistic charge--dipole model for Ag and Au nanoparticles directly to first-principles induced-density profiles. The calibration includes the short-range kinetic and exchange-correlation contribution to the hardness kernel that a purely electrostatic model omits, reducing the induced-density-profile error by about a factor of 2.5. A single averaged parameter set per element predicts clusters withheld from the fit and transfers without adjustment to Ag/Au core--shell, alloyed, and elongated particles. A continuous fast multipole implementation of the model scales almost linearly with the number of atoms and reaches several-million-atom particles, far beyond the reach of first-principles methods.
\end{abstract}

\section{Introduction}

External electric fields redistribute the conduction electrons of a metallic nanoparticle to screen the perturbation.
This redistribution is recorded by the induced charge density $\Delta\rho(\mathbf{r})$ with full spatial resolution: where charge accumulates and depletes, how deeply screening penetrates into the interior, and how the response varies across facets, edges, and vertices.
First-principles calculations show that even clusters of tens of atoms expel the interior field almost completely.\cite{sinharoy2020,sinharoy2023}
Some observables of this density can be measured: molecular-beam electric deflection gives the static polarizability of size-selected clusters,\cite{heiles2014deflection} and off-axis electron holography reconstructs the external electrostatic potential around nanostructures.\cite{zheng2020holography,zheng2023tomography}
The same atomic-scale features underlie dynamic nanoplasmonics such as near-field hot spots, surface-enhanced Raman scattering, and hot-carrier generation,\cite{sers_review_schlucker,plasmon_hot_carriers_brongersma,lspr_review_maier} which require a frequency-dependent treatment not attempted here.
Resolving this site-dependent response requires a model that represents $\Delta\rho(\mathbf{r})$ atom by atom.

Density functional theory (DFT) provides a parameter-free route to $\Delta\rho(\mathbf{r})$, but conventional Kohn--Sham implementations scale as $\mathcal{O}(N^3)$ with the number of atoms $N$, restricting routine induced-density calculations to at most a few hundred to $\sim\!10^3$ atoms.\cite{ksdft_scaling_review,metals_linear_scaling_limits}
The 20--50~nm particles probed in single-particle holography\cite{zheng2020holography,zheng2023tomography} and optical spectroscopy\cite{sonnichsen2002damping,olson2015singleparticle} contain $10^5$--$10^6$ atoms,\cite{au_lattice_constant} an orders-of-magnitude gap.
Continuum electrodynamics, through the boundary-element method (BEM), finite-difference time-domain (FDTD), and discrete-dipole-approximation (DDA) algorithms, handles these sizes routinely.
These methods represent the particle by a bulk dielectric response and a smooth surface, or in the DDA by a grid of point dipoles.
Facets, edges, and atomic-scale roughness are therefore absent by construction, and surface or nonlocal corrections require further fitted parameters.\cite{bem_plasmonics_review,bem_garcia_de_abajo,bem_surface_corrections_esteban,taflove2005fdtd,draine1994dda,kelly2003dda,nonlocal_drude_raza,mortensen2021review}
Jellium models and real-time time-dependent DFT (TDDFT) retain quantum effects but either discard morphology or remain prohibitive at the $10^4$--$10^6$-atom scale.\cite{kuisma2015tddft,jellium_tddft_yan,tddft_clusters_review}
A model that is calibrated to finite-particle DFT induced densities, tested outside the fitting set, and retains explicit atomic morphology at experimentally relevant sizes has not yet been established.

Atomistic polarizable models such as charge-equilibration, Drude-oscillator, and charge--dipole schemes\cite{qeq_eem_review,mortier1986eem,drude_polarizable_review,applequist1972adi,thole1981,olson1978monopoledipole,charge_dipole_mayer,mayer2008chargedipole,mayerschatz2009agclusters,atomistic_large_scale_jensen} also reach these sizes while treating atoms explicitly.
Recent fluctuating-charge and fluctuating-dipole formulations for nanoplasmonics, including bimetallic particles, have handled $10^4$--$10^6$ atoms.\cite{chen2015cddim,giovannini2019wfq,giovannini2022wfqfmu,nicoli2023bimetallic,nicoli2024colloidal,verstraelen2013acks2,lafiosca2021million}
Related constant-potential electrode models reproduce metallic screening at rough, atomically structured surfaces.\cite{siepmann1995electrode,scalfi2021electrode}
The accuracy of these models depends on how they are calibrated, and most parameterizations use integrated observables such as bulk dielectric constants, polarizabilities, or induced dipole moments.\cite{drude_polarizable_review,drude_jensen_metal_clusters}
Such quantities constrain the spatial distribution only weakly, because different induced-density profiles can yield the same total dipole,\cite{Bodrenko2012,atomistic_transferability_issues,charge_models_limitations} so a model fit to dipoles may place the screening charge incorrectly.
For periodic silver slabs, Bodrenko et al.\ showed that a continuum of charge--dipole parameterizations reproduces the slab polarizability while describing the induced density inside the slab incorrectly, and introduced an integrated squared-density measure to remove that ambiguity.\cite{Bodrenko2012}
Equivariant machine-learning models provide an emerging route to the field-induced electron density at the thousand-atom scale,\cite{lewis2023densityresponse,rossi2025vfield} but they require system-specific training data.
Here we pursue a transferable, physically parameterized model fit to that density.
The induced-density profile has been computed from first principles for noble-metal clusters\cite{sinharoy2020,sinharoy2023,ma2015agcluster} and used to validate electrostatic models for periodic slabs,\cite{Bodrenko2012,Bodrenko2013} but, to our knowledge, it has not been adopted as the primary calibration target for a transferable atomistic model of finite Ag/Au nanoparticles across size, morphology, and bimetallic composition.
That density-resolved calibration, and its predictive range, is what we test here.

Accuracy is also limited by how the induced distributions interact.
In the charge--dipole model of Bodrenko et al.,\cite{Bodrenko2012} the induced Gaussian distributions interact purely electrostatically.
The static response of a metal, however, is governed by the hardness kernel of DFT, which adds to the long-range Coulomb term a short-range kinetic and exchange--correlation (KXC) contribution.\cite{Bodrenko2013}
Omitting it limits how accurately a purely electrostatic model can reproduce the induced charge density, even when the dipole is fit well.
For periodic silver slabs, Bodrenko and Della Sala\cite{Bodrenko2013} showed that approximating this term by overlap integrals of the basis functions, with one extra coefficient per channel, substantially improves the induced-density profile while preserving the model's favourable scaling.
In the local infinite-jellium limit, the same form reduces to a constant KXC hardness.
That correction has not previously been parameterized or tested for finite metal clusters or for chemically heterogeneous (bimetallic) systems.

Here we calibrate the KXC-corrected charge--dipole model directly against finite-field DFT induced charge densities of Ag and Au clusters.
The charge--dipole engine and the overlap form of the KXC correction are established.\cite{Bodrenko2012,Bodrenko2013} What we add is their calibration and testing for finite nanoparticles rather than periodic slabs, and the demonstration that one parameter set per element transfers across size, morphology and composition to give an atomically structured screening layer that persists to the conductor limit.
We show that the KXC correction is necessary within this model class for quantitative agreement with the DFT finite-width surface screening layer.
The resulting element-averaged parameters predict clusters withheld from the fit and are tested on bimetallic Ag/Au particles spanning core--shell, alloyed, and elongated geometries.
The calibrated density also improves the DFT-derived polarizabilities and external potentials that can be compared with future experiments.
When applied unchanged to larger fcc particles, the model recovers conductor-like integrated screening at nanometre sizes while still resolving the facet, edge, and vertex charge that smooth continuum descriptions omit.
Continuous fast multipole evaluation of the Coulomb interactions\cite{Lazarski2015} makes nanoparticles of up to $\sim\!1.5\times10^{7}$ atoms efficiently tractable on a single CPU workstation, comparable to the largest fully atomistic classical-plasmonics simulations.\cite{lafiosca2021million}
This provides a density-calibrated, atomically resolved route to the static screening response of metal nanoparticles at the sizes relevant to single-particle holography and plasmonic experiments.

\section{Results}
\label{sec:results}

\subsection{Model and calibration strategy}
\label{sec:glance}
The model places the induced charge density on the atoms.
Each atom $i$ carries an induced Gaussian charge $q_i$ and an induced Gaussian dipole $\bm\mu_i$.
Two element-specific widths, $R_q$ and $R_p$, are the Gaussian radii that set the spatial extent of the induced charge and dipole clouds (distinct from the surface-layer width $w_r$ introduced later).
The total induced dipole is $\sum_i(q_i\mathbf R_i+\bm\mu_i)$.
The charges and dipoles respond linearly to an applied field through a quadratic interaction energy whose minimization is a single linear solve (Methods).

In the uncorrected periodic charge--dipole electrostatic model (PCDEM),\cite{Bodrenko2012} these induced distributions interact only through the Coulomb kernel.
The kinetic-exchange-correlation (KXC) correction keeps the same degrees of freedom but adds a short-range overlap term to the interaction.
Two further per-element coefficients, the couplings $k_q$ and $k_p$, set the strength of this short-range correction in the charge and dipole channels.
Setting $k_q=k_p=0$ recovers PCDEM.
Four parameters per element, $(R_q,R_p,k_q,k_p)$, thus define the static response.

The parameters are fixed by matching the model induced density to finite-field DFT references. The fitting objective is the one-dimensional integrated-square-density (1D-ISD) error $\Delta_{\text{1D-ISD}}$: the relative root-mean-square (RMS) deviation, averaged over the three field directions, of the plane-integrated induced-density profile $P(u)=\iint\Delta\rho\,\mathrm{d}v\,\mathrm{d}w$ from its DFT counterpart (defined in Methods). It scores the spatial fidelity of the screening charge, where the charge sits and how wide the surface layer is, not merely its integral. The relative dipole error $\Delta_{\mathrm{dip}}$ is monitored as an independent global-response check and is never part of the objective.

This distinction matters because the dipole is only the first moment of $\Delta\rho$: different screening profiles can share the same dipole,\cite{atomistic_transferability_issues,charge_models_limitations} so a dipole-fitted model can place the surface charge incorrectly while appearing accurate, whereas calibrating to the profile removes this ambiguity. We first test whether the KXC correction fixes the DFT screening profile for the icosahedral clusters \ce{Ag55} and \ce{Au55}, where finite-field DFT references are available.

\subsection{Density calibration reveals the missing surface response}
\label{sec:elemental}

For icosahedral \ce{Ag55} and \ce{Au55} the optimized parameters are $(R_q,R_p,k_q,k_p)=(3.48,2.90,36.4,34.8)$ and $(3.44,2.82,48.4,36.4)$, respectively (widths in $a_0$, couplings in $a_0^2$).
The uncorrected PCDEM model ($k_q=k_p=0$) optimizes to $(3.14,2.36)$ for Ag and $(3.00,2.30)$ for Au (full table in the Supplementary Material).
The KXC correction lowers the 1D-ISD profile error from $19.1\%$ to $7.6\%$ for Ag and from $21.6\%$ to $8.8\%$ for Au, by a factor of about $2.5$, while the dipole error drops by an order of magnitude, from $5.2\%$ to $0.4\%$ for Ag and from $7.2\%$ to $0.7\%$ for Au.
Both improvements are obtained with the same number of fields and reference data.
The only change is the inclusion of the short-range KXC term in the hardness matrix.
The per-direction errors agree to within $0.1$ percentage point, as the icosahedral symmetry requires (Supplementary Material). These are per-cluster optima, reported here to isolate what the correction does. All subsequent transfer and observable results instead use a single element-averaged parameter set (Methods). Under that set the \ce{Ag55} and \ce{Au55} profile errors are $11.5\%$ and $10.4\%$ and the dipole errors $3.0\%$ and $4.0\%$ (Supplementary Material), and the derived static observables they yield are reported below.

\begin{figure*}[htbp]
\centering
\includegraphics[width=\textwidth]{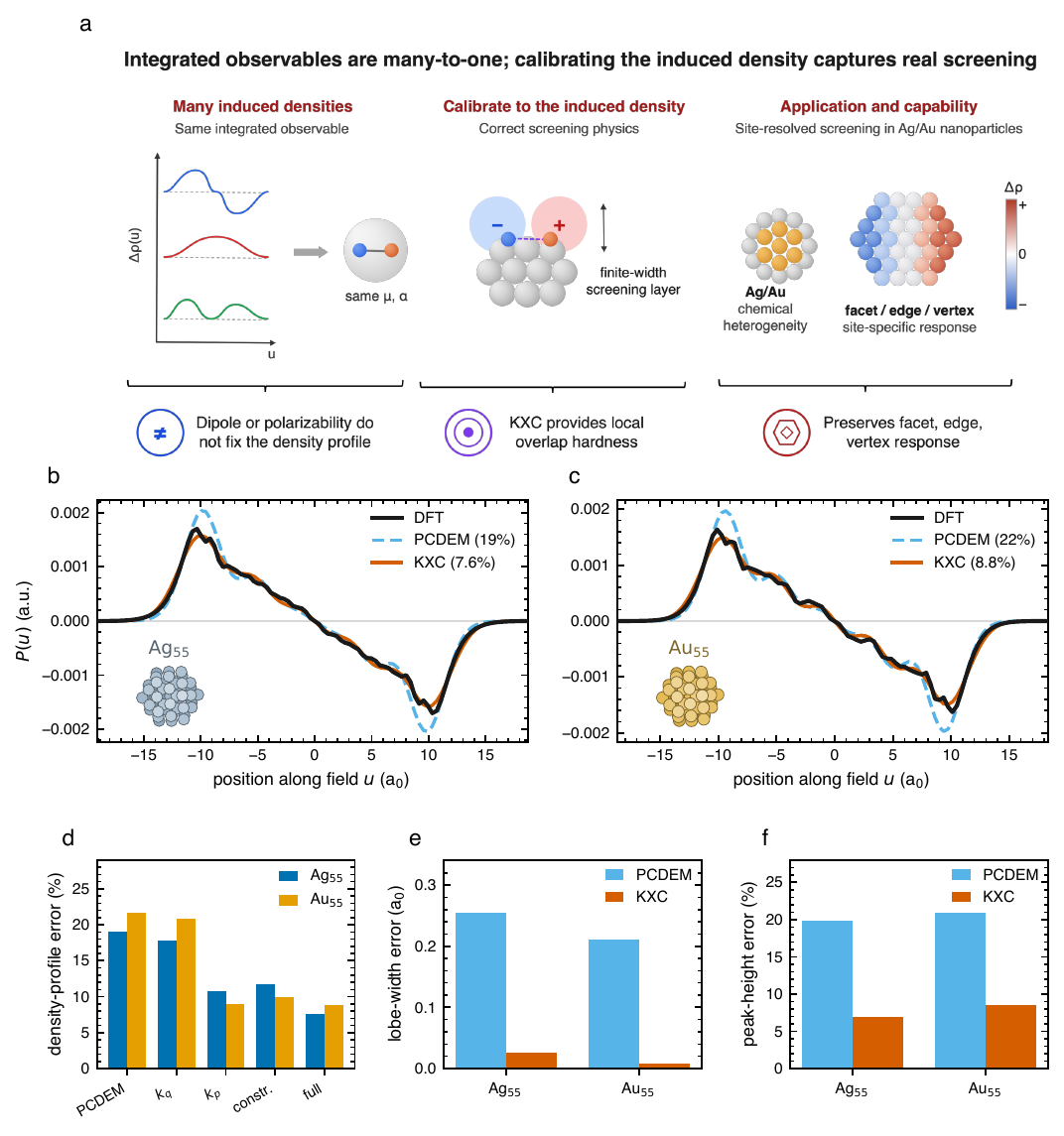}
\caption{Concept, validation, and mechanism of the KXC correction. (a) Schematic summary of density-profile calibration and its use for site-resolved Ag/Au nanoparticle screening. (b,c) $P(u)$ of icosahedral \ce{Ag55} and \ce{Au55} under a field of $10^{-4}$~a.u.\ along $u$: DFT reference (black), PCDEM (dashed blue), and KXC (solid vermilion). Percentages are 1D-ISD profile errors. (d) Channel decomposition of the direction-averaged 1D-ISD error. (e,f) Lobe-width and peak-height errors of the induced surface-charge layer.}
\label{fig:profiles}
\end{figure*}

Figure~\ref{fig:profiles} shows the origin of this improvement directly.
It compares the plane-integrated induced-density profile $P(u)$ of the DFT reference with those of the two models at their optima, for \ce{Ag55} and \ce{Au55}.
The profile is antisymmetric, with charge of opposite sign accumulating on the two sides of the cluster along the field, and its first moment is the induced dipole.
The uncorrected model systematically overshoots the two surface peaks, the signature of fitting a peaked response with a purely electrostatic kernel.
The KXC-corrected model follows the reference closely across the whole profile, peaks and interior alike.
A positive dipole coupling $k_p$ reduces the model polarizability relative to the electrostatic-only value [Eq.~\eqref{eq:alpha_kxc}], which suppresses the overshoot.
The channel decomposition below confirms that this dipole-channel term carries most of the correction.

At the profile optimum the dipole error is already $\le0.7\%$, so fitting the spatial profile leaves the integrated response essentially unchanged.
By contrast, fitting the dipole alone gives a worse profile (Supplementary Material).
The profile is therefore the binding constraint, and the gain lies in the spatial form of the surface charge rather than in the dipole.
We localize it next.

\subsection{KXC corrects the finite-width screening layer}
\label{sec:ablation}

To check that the improvement reflects the physics of the correction and not merely two extra fit parameters, we enabled the charge and dipole channels separately, reoptimizing the allowed parameters against the 1D-ISD error in each case.
Figure~\ref{fig:profiles}(d) reports the result for \ce{Ag55} and \ce{Au55} (full channel-decomposition values in the Supplementary Material).
The dipole-channel coupling $k_p$ alone recovers most of the improvement: the profile error drops from $19.1\%$ (PCDEM) to $10.7\%$ for Ag and from $21.6\%$ to $9.0\%$ for Au, whereas the charge-channel coupling $k_q$ alone barely changes it ($17.8\%$ and $20.9\%$).
The correction therefore acts mainly on the dipolar response.
A positive $k_p$ suppresses the over-screening of the induced dipole [Eq.~\eqref{eq:alpha_kxc}].
The charge channel and the freedom to set the two widths independently add a secondary refinement, reaching the final $7.6\%$ (Ag) and $8.8\%$ (Au).
A constrained variant with $R_q=R_p$ and $k_q=k_p$, the restriction adopted for periodic slabs\cite{Bodrenko2013}, reaches $11.8\%$ (Ag) and $10.0\%$ (Au), better than either single channel but short of the unrestricted fit, confirming that for finite clusters the four independent parameters are warranted.

\label{sec:descriptors}
The 1D-ISD error is an aggregate norm.
Decomposing the same profiles into surface-charge-lobe descriptors (Supplementary Material) localizes what the correction does.
Both models place the induced-charge centroid accurately (within $\sim\!0.1~a_0$ of DFT), so the position of the screening charge is already set by the electrostatic kernel.
KXC instead fixes its spatial extent [Fig.~\ref{fig:profiles}(e,f)].
It reduces the lobe-width error from $\sim\!0.25~a_0$ (PCDEM) to $\le0.03~a_0$ and the peak-height error from $\sim\!20\%$ to $7$--$9\%$.
The uncorrected model reproduces a peaked surface response only by making each lobe too narrow and too tall.
The short-range KXC term broadens and lowers it to match DFT.
These centroid and width descriptors are the static analogue of the induced-charge position and spatial extent that enter Feibelman-type surface-response theories of nanophotonics.
The static centroid of the field-induced surface charge is the classical image-plane position of Lang and Kohn.\cite{feibelman1982,langkohn1973,mortensen2021review,mortensen2021surfaceresponse,goncalves2021dparam,chen2025dparams}
A full three-dimensional comparison (Supplementary Material) is consistent: KXC raises the collective response amplitude $A_{\rm opt}$ from $\sim\!0.6$ toward unity and the density-pattern correlation $C_\rho$ from $\sim\!0.73$ to $\sim\!0.82$ for both \ce{Ag55} and \ce{Au55}.
The remaining difference is atomic-scale density texture that the smooth Gaussian basis cannot represent and that averages out of the integrated observables.

The validation proceeds in four steps.
We first fit \ce{Ag55} and \ce{Au55} to isolate the KXC mechanism.
We then average parameters over compact monometallic clusters.
We test prediction by leave-one-cluster-out validation.
Finally, we apply the same fixed parameters to bimetallic particles and larger fcc spheres.
We next test whether the element-specific parameters transfer across cluster size and morphology.

\subsection{Element-averaged parameters predict unseen clusters}
\label{sec:multicluster}

The parameters above came from a single icosahedral cluster per element.
To test whether they are transferable element-specific quantities rather than fits tied to one geometry, we repeated the four-parameter KXC fit independently for additional metallic Ag and Au clusters.
The set spans decahedral, truncated-cube, truncated-octahedron, octahedral, and icosahedral morphologies, with sizes of 55--101 atoms.
Each cluster has a tight-convergence DFT reference and the same coarse-to-fine parameter search.

Figure~\ref{fig:cluster_params}(a--d) displays the per-cluster optima against cluster size (full values in the Supplementary Material).
For a given element the four parameters concentrate around common values.
The widths agree to within $\sim\!0.2~a_0$, apart from one truncated-cube cluster with a smaller $R_q$, and the couplings vary by $\sim\!10$--$20\%$.
No systematic size dependence is apparent over the 55--101-atom range.
This supports using element-averaged static-response parameters over the tested compact, neutral Ag/Au cluster regime.
We therefore define one element-averaged parameter set per element as the average over the clusters: $(R_q,R_p,k_q,k_p)=(3.36,2.93,44.6,34.6)$ for Ag and $(3.35,2.82,49.1,32.5)$ for Au.
The uncorrected model is averaged in the same way from its own per-cluster optima, giving PCDEM widths $(R_q,R_p)=(3.06,2.49)$ for Ag and $(2.92,2.37)$ for Au.
These PCDEM widths, not the KXC widths with the couplings set to zero, are used in the transferability tests.
These averages coincide, to within the scatter, with the single-cluster icosahedral values, so the icosahedral fit already captured the element-specific physics.

\begin{figure}[htbp]
\centering
\includegraphics[width=\columnwidth]{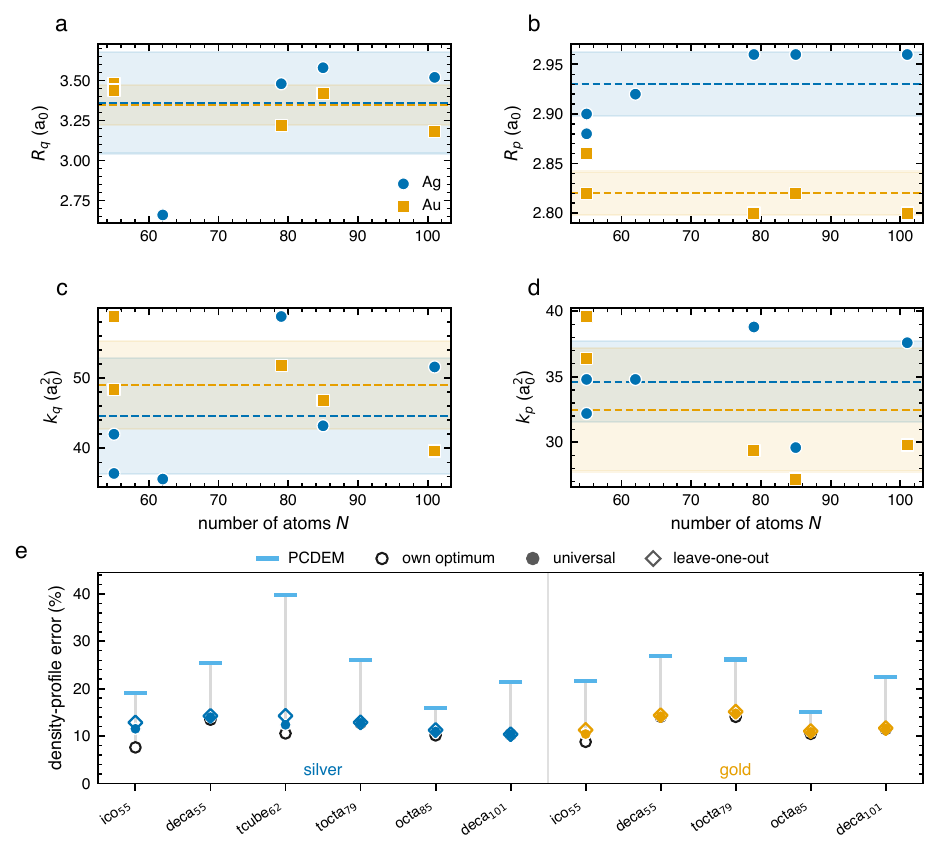}
\caption{One element-averaged parameter set per element. (a--d) KXC parameters $R_q$, $R_p$, $k_q$, $k_p$ from independent per-cluster fits versus cluster size $N$, for silver (circles) and gold (squares). Dashed lines and shaded bands are the element mean $\pm$ one standard deviation. (e) Direction-averaged density-profile error of each cluster for the uncorrected model (PCDEM, caps), the cluster's own optimum (open circles), the element-averaged set (filled circles), and the leave-one-out prediction in which the held-out cluster contributes nothing to the parameters (open diamonds). Full tables in the Supplementary Material.}
\label{fig:cluster_params}
\end{figure}

We next tested one element-averaged parameter set on every cluster, with no per-cluster adjustment, and compared the result with the independent DFT reference [Fig.~\ref{fig:cluster_params}(e), with full tables in the Supplementary Material].
Across the set, the element-averaged parameters reproduce the induced-density profile with $\Delta_{\text{1D-ISD}}=12.2\%$, only $0.9$ percentage points above the per-cluster optima ($11.3\%$).
The spatial response is therefore transferable at near-optimal accuracy across the tested sizes and morphologies.
The dipole error is more cluster-specific: $\Delta_{\mathrm{dip}}=1.9\%$ for the element-averaged set against $0.7\%$ at the individual optima.
This mirrors the shallower dependence of the integrated dipole on the parameters seen in the elemental fits.
For every cluster the KXC model is more accurate than the uncorrected one (Supplementary Material).

\label{sec:loo}
Applying the element-averaged parameters back to the clusters they were averaged from is not a blind test.
To assess predictive transfer we performed a leave-one-cluster-out (LOO) cross-validation.
For each metallic cluster the element average was recomputed from the other clusters only, and the held-out cluster was evaluated at those parameters, with no information from it entering its own prediction [open diamonds in Fig.~\ref{fig:cluster_params}(e), with full table in the Supplementary Material].
The LOO 1D-ISD error (mean $12.7\%$) tracks the full-average error (mean $12.2\%$) closely.
Withholding a cluster degrades its prediction by $0.5$ percentage points on average, and by at most $1.9$ percentage points for the truncated-cube cluster, the main outlier of Fig.~\ref{fig:cluster_params}.
Every LOO prediction remains more accurate than the uncorrected model (mean $23.6\%$).
The element parameters therefore predict the induced-density profile of clusters they were not fitted to.
This motivates applying the same element-averaged parameters to chemically heterogeneous particles within controlled bimetallic tests.

\subsection{Transfer to bimetallic Ag/Au systems}

The leave-one-out analysis selects the element-averaged parameters as the predictive choice, so we apply them, held fixed, to chemically heterogeneous particles for both the uncorrected (PCDEM) and corrected (KXC) models.
A single-cluster fit can transfer marginally better to a target of its own morphology, but the element-averaged set is more accurate across morphologies on average, so we use it in the remainder of the work.
We test six bimetallic systems not included in the fit. Four are the 55-atom core--shell cluster \ce{Au13}@\ce{Ag42} (13 gold atoms in the core, 42 silver atoms in the surface shell) in distinct morphologies: icosahedral, cuboctahedral, decahedral, and a low-symmetry distorted cuboctahedron. The other two are mixed systems in which both elements reach the surface: a compact 55-atom alloy of 27 gold and 28 silver atoms, and a 60-atom silver-rich nanorod (48 Ag, 12 Au).
Every silver atom is assigned the silver parameters and every gold atom the gold parameters.
The optical response of Ag/Au nanoalloys and core--shell particles has previously been studied by first-principles and atomistic methods;\cite{lopezlozano2013agau,nicoli2023bimetallic} here we test whether the static element-averaged parameters transfer to these compositions.

\begin{figure}[htbp]
\centering
\includegraphics[width=\columnwidth]{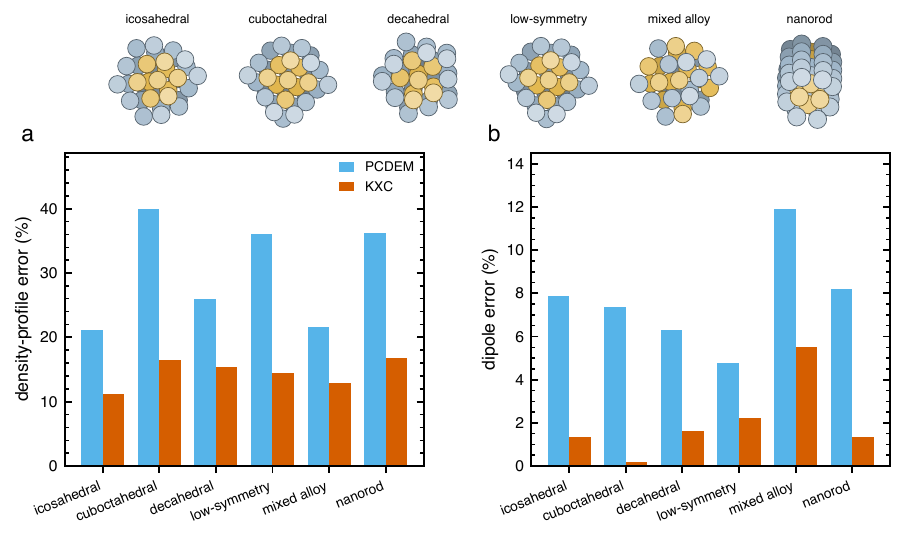}
\caption{Transfer of the element-averaged parameters to six bimetallic Ag/Au systems. Top: the four \ce{Au13}@\ce{Ag42} core--shell morphologies (cut open: gold core, silver shell) and the two mixed systems in which gold is exposed (a compact \ce{Au27Ag28} alloy and an \ce{Ag48Au12} nanorod). (a) Density-profile (1D-ISD) error and (b) dipole error of the uncorrected (PCDEM) and corrected (KXC) models, both with the element-averaged parameters held fixed. The first four systems are the core--shell morphologies and the last two the mixed alloy and nanorod. The KXC correction roughly halves the profile error for every system. Full per-direction tables in the Supplementary Material.}
\label{fig:transferability}
\end{figure}

Figure~\ref{fig:transferability} reports the transfer of the element-averaged parameters to the six systems.
The uncorrected model is inaccurate on every system (direction-averaged $\Delta_{\text{1D-ISD}}=21$--$40\%$), and the KXC correction roughly halves it, to $11$--$17\%$.
The corrected error, averaged over the four core--shell morphologies, is $14.4\%$.
The faceted shells (cuboctahedral $16.5\%$, low-symmetry $14.5\%$) carry a larger residual than the icosahedral ($11.2\%$).
This is consistent with the (100) facets, edges, and vertices they expose screening the field differently from site to site, which a single isotropic width per element may not fully capture.
First-principles studies likewise find facet-dependent surface response on noble metals, including different screening on (111)- and (100)-type facets,\cite{echarri2021facets} crystal-face-dependent work functions,\cite{tran2019workfunction} and a coordination-sensitive induced response.\cite{rossi2020hotcarriers}
The dipole error is $0.2$--$2.2\%$ across the core--shell morphologies.

In the core--shell clusters gold is confined to the buried 13-atom core, so the silver shell carries $61$--$67\%$ of the induced dipole and the transfer is largely insensitive to the gold parameters. Replacing the gold average by an alternative metallic average shifts the 1D-ISD error by at most a few tenths of a percentage point (Supplementary Material), a useful robustness property for core--shell architectures.
The two mixed systems remove this shielding, with gold at the surface throughout the alloy and along the rod, so both elements contribute directly to the tested surface screening layer.
The element-averaged set still transfers. The corrected profile error is $12.8\%$ for the alloy, as accurate as the core--shell average, and $16.8\%$ for the nanorod, comparable to the faceted core--shell shells.
The dipole error remains $1.3\%$ for the rod and rises to $5.5\%$ for the alloy, where the exposed gold makes the global response harder to reproduce, still well below the $11.9\%$ of the uncorrected model.
Together they support transfer of one element-averaged parameter set across the bimetallic geometries examined here.
The residual is consistent with the present isotropic, site-independent parameters, which assign one width and one coupling to every atom of an element.
The site dependence of the screening is already encoded in the geometry. With these same parameters the model reproduces the facet-, edge-, and vertex-resolved surface charge from the atomic positions alone (next section).
An explicit coordination-dependent parameterization could be tested in future, but the present data indicate that geometry already captures the dominant site structure.
Having shown transfer beyond the elemental clusters, we next examine whether the calibrated density also improves the physical observables derived from the response.

\subsection{Static observables from the calibrated density}
\label{sec:bridge}

With the element-averaged parameters held fixed, the model yields static observables that can be compared with experiment in favourable cases.

\begin{figure*}[htbp]
\centering
\includegraphics[width=\textwidth]{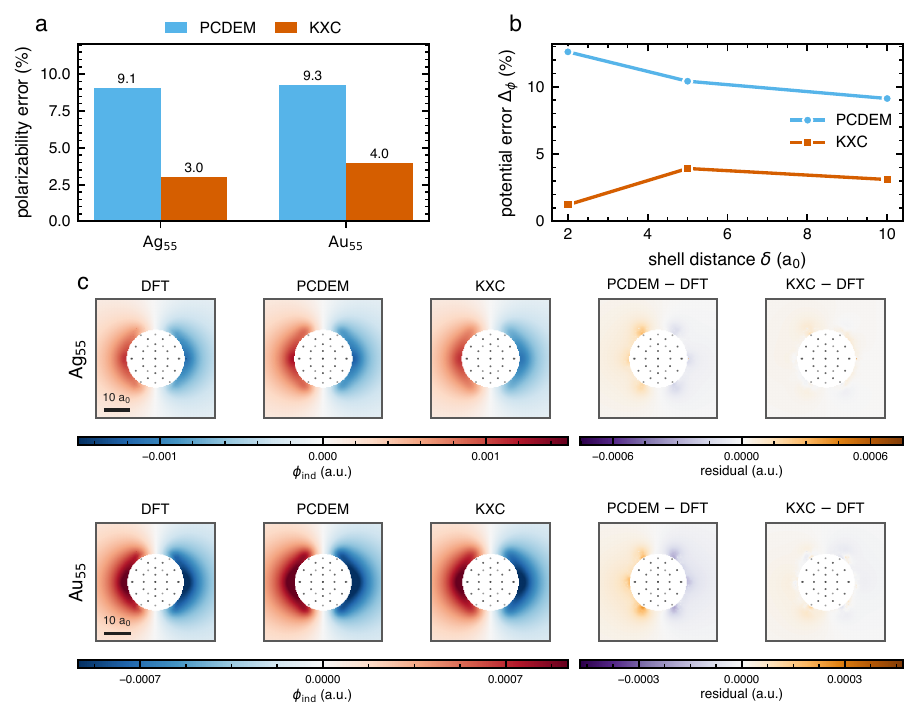}
\caption{Static observables from the calibrated density. (a) Relative error of the static polarizability $\alpha=\mu/E_0$ for \ce{Ag55} and \ce{Au55} (KXC $3$--$4\%$ vs $\sim\!9\%$ for PCDEM). (b) Relative error of the induced electrostatic potential on shells at distance $\delta$ beyond the ionic surface of \ce{Ag55}. (c) External induced potential $\phi_{\rm ind}$ in the central plane of \ce{Ag55} (top row) and \ce{Au55} (bottom row), field along the horizontal axis: DFT, the two models (shared colour scale), and the model$-$DFT residuals (shared residual scale). Only the region outside the ionic surface is shown (atoms as dots).}
\label{fig:observables}
\end{figure*}

The static polarizability is the linear dipole response, $\alpha=\mu/E_0$.
For the icosahedral references the DFT values are $1847~a_0^3$ for \ce{Ag55} ($33.6~a_0^3$ per atom) and $1603~a_0^3$ for \ce{Au55} ($29.2~a_0^3$ per atom).
The KXC model reproduces them to within $4\%$ ($3.0\%$ and $4.0\%$), against about $9\%$ for the uncorrected model [Fig.~\ref{fig:observables}(a)].
As an integrated quantity the polarizability does not constrain the density profile, so it serves here as a consistency check rather than a fitting target.
Molecular-beam electric deflection offers a direct measurement route\cite{heiles2014deflection,knight1985polarizability} and is well established for alkali and other clusters. For free Ag and Au clusters in this size range we are not aware of a direct static-polarizability measurement. The closest experimental data are the optical-absorption spectra of size-selected neutral Ag clusters,\cite{yu2018agclusters} from which a static polarizability can in principle be recovered through the oscillator-strength sum rules, so a comparison with experiment is at present indirect.

The induced electrostatic potential outside the particle, $\phi_{\rm ind}(\mathbf r)=\int\Delta\rho(\mathbf r')/|\mathbf r-\mathbf r'|\,d^3r'$, is the quantity reconstructed by off-axis electron holography and holographic tomography,\cite{zheng2020holography,zheng2023tomography} which has been applied to the electrostatic potential of metal nanoparticles\cite{ozsoy2016holography} and is reviewed as a quantitative probe of nanoscale potentials and charges.\cite{mccartney2019holographyreview}
On shells $2$--$10~a_0$ beyond the ionic surface the KXC model gives a potential error of $1.2$--$4.4\%$ with near-unit correlation, several times smaller than the $9$--$13\%$ of the uncorrected model [Fig.~\ref{fig:observables}(b)].
The plane maps of Fig.~\ref{fig:observables}(c), for both \ce{Ag55} and \ce{Au55}, show the model$-$DFT residual to be several times smaller for the corrected model throughout the near-surface region.
The consequence for the field at a molecule-scale standoff is quantified in the Supplementary Material.
Because holography can resolve the electrostatic signature of single elementary charges on individual nanoparticles,\cite{gatel2013counting,aso2022science} the external potential is a possible experimental observable for future tests of the model.
Here, however, it is benchmarked only against DFT.

These comparisons confirm that the calibrated density reproduces the derived observables against DFT. We next take the same static model to nanoparticle sizes beyond the reach of DFT.

\subsection{Atomistic-to-continuum crossover}
\label{sec:largeparticle}

\begin{figure*}[htbp]
\centering
\includegraphics[width=0.92\textwidth]{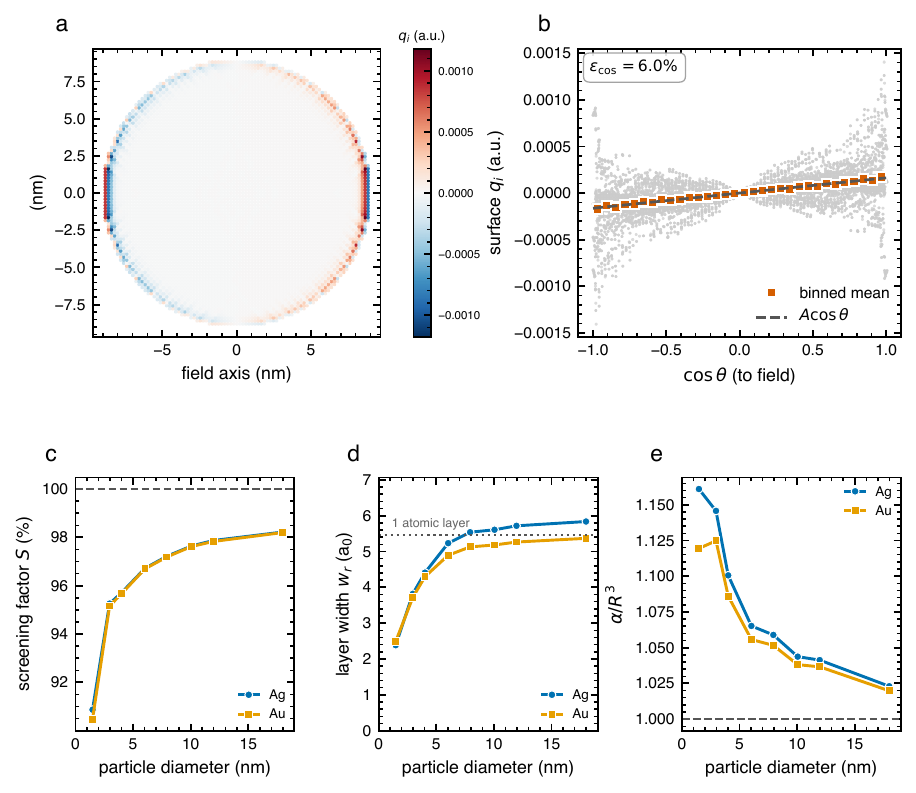}
\caption{Atomistic-to-continuum crossover of static screening. (a) Cross-section of the 18~nm sphere ($1.8\times10^{5}$ atoms) in a uniform static field, atoms coloured by induced charge $q_i$. (b) Surface-atom charge versus $\cos\theta$ to the field with the least-squares conductor law $A\cos\theta$. (c--e) Size dependence: (c) interior screening factor $S$ rising toward the conductor limit, (d) screening-layer width $w_r$ saturating near one nearest-neighbour distance (dotted), and (e) polarizability ratio $\alpha/R^3\to1$.}
\label{fig:large_particle}
\end{figure*}

With the element-averaged parameters held fixed, the model gives the static response of nanoparticles across the nanometre size range (Fig.~\ref{fig:large_particle}).
For an 18~nm fcc-silver sphere ($1.8\times10^{5}$ atoms), the largest we treat, in a uniform static field, the induced charge accumulates in the field-facing surface region [Fig.~\ref{fig:large_particle}(a)].
When binned by polar angle $\theta$ to the field, the surface charge follows the classical conductor law $\sigma(\theta)\propto\cos\theta$ for a sphere in a uniform field\cite{jackson_electrodynamics} [Fig.~\ref{fig:large_particle}(b)].
We quantify the agreement by the conductor-law residual $\epsilon_{\cos}$, the relative deviation of the angularly binned surface charge from a least-squares $A\cos\theta$ fit (Methods), which is $6\%$ at 18~nm.
From 6~nm upward the residual decreases with size, from $85\%$ at 6~nm to $9\%$ at 12~nm and $6\%$ at 18~nm for Ag.
Below 6~nm it is not monotonic, because the binned surface charge is dominated by the discrete termination of each model sphere (Supplementary Material).
The dipolar far field of this surface charge is, in principle, accessible to electron holography.\cite{zheng2020holography,zheng2023tomography}
The particle size is comparable to those for which holographic instruments have demonstrated single-charge sensitivity.\cite{gatel2013counting,aso2022science}
Such a reconstruction of field-induced surface charge has not yet been performed, and the present results are benchmarked only against DFT.

The integrated response approaches the macroscopic conductor more regularly than the local conductor-law residual.
The interior screening factor $S$ (Methods) is $94\%$ for \ce{Ag55} and $90\%$ for \ce{Au55} at the cluster scale.
These icosahedral cluster values are not expected to lie exactly on the fcc-sphere trend, because the shapes and core masks differ.
They are of the same order as the $\sim\!96\%$ reported for noble-metal clusters from first principles,\cite{sinharoy2020,sinharoy2023} but the estimators and core definitions differ, so the numbers should not be compared point by point (Supplementary Material).
The factor rises to about $97\%$ by 6~nm and approaches $98\%$ at 12--18~nm [Fig.~\ref{fig:large_particle}(c)].
The polarizability ratio $\alpha/R^3$ descends toward the conductor-sphere value of unity,\cite{jackson_electrodynamics,mie_theory_bohren_huffman} reaching $1.04$ at 12~nm and $1.02$ at 18~nm [Fig.~\ref{fig:large_particle}(e)].
The screening layer itself does not sharpen to a continuum surface.
We measure its thickness by the charge-weighted radial spread of the induced charge, $w_r$ (Methods), which rises with size and then saturates at about one atomic nearest-neighbour distance rather than growing in proportion to the particle radius [Fig.~\ref{fig:large_particle}(d)].
This width includes the atomistic surface corrugation and the finite extent of the atom-centred basis.
It should not be interpreted as a frequency-dependent Feibelman width.
The static response thus crosses over from molecular clusters to conductor-like integrated behaviour while retaining a finite-width, atomically structured screening layer.

\subsection{Facet, edge, and vertex screening}
\label{sec:heterogeneity}

\begin{figure*}[htbp]
\centering
\includegraphics[width=\textwidth]{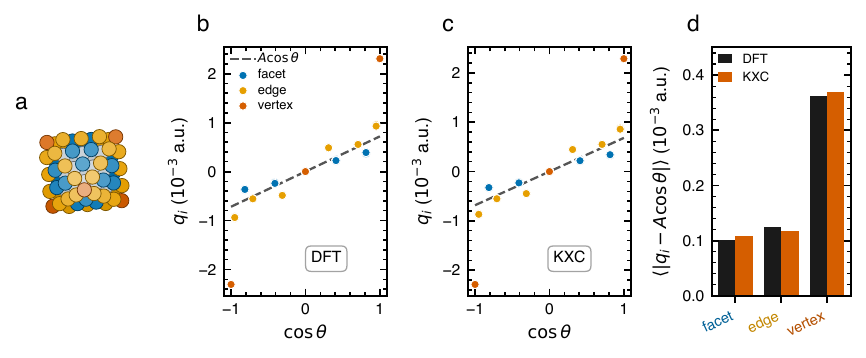}
\caption{Screening beyond the smooth-surface picture. (a) Octahedral \ce{Ag85} with surface sites coloured by type: facets (blue), edges (gold), vertices (vermilion). (b,c) Per-atom induced charge $q_i$ versus $\cos\theta$ to the field for (b) the DFT reference and (c) the KXC model, with sites coloured as in (a). The dashed line is the smooth continuum law $A\cos\theta$. Vertex atoms depart most strongly from the line. (d) Mean departure $|q_i-A\cos\theta|$ by site type, averaged over the faceted clusters, for DFT and KXC: vertices are several times more heterogeneous than facets (on average $3.5\times$), and the site-independent KXC model reproduces the trend.}
\label{fig:heterogeneity}
\end{figure*}

A continuum description assigns the surface charge a single smooth angular profile $\sigma(\theta)\propto\cos\theta$, whereas the atomic structure of a real faceted particle makes the local screening charge depend on the coordination of each surface site.
We quantify this on the faceted clusters, whose DFT references resolve it directly.
For each surface atom we obtain its net induced charge by a nearest-atom (Voronoi) partition of the induced density, classify it as a facet, edge, or vertex site from its coordination number, and measure the departure of its charge from the smooth conductor law, $\delta q_i=q_i-A\cos\theta_i$.
This departure is site dependent, with vertex atoms deviating from the $\cos\theta$ law on average $3.5$ times as much as facet atoms (a factor of $2.4$ to $4.7$ across the faceted clusters) [Fig.~\ref{fig:heterogeneity}], an atomistic feature absent by construction from any geometrically smooth surface.
The KXC model, although its parameters are isotropic and site independent (one width and one coupling per element), reproduces this heterogeneity from the geometry alone.
These faceted clusters are among those from which the element-averaged parameters were built, so this is not a parameter-blind transfer test.
It is a representation check, because the model was calibrated only to the direction-averaged one-dimensional density profile and never to per-atom charges or to the site-resolved residuals examined here.
Because the raw per-atom charge is dominated by the global $\cos\theta$ trend, we compare the heterogeneity on the residuals themselves.
After subtracting the fitted conductor law, the relative error of the site-resolved model residuals is $5$--$7\%$ across the faceted clusters.
This is three to five times smaller than the uncorrected model's residual error of $20$--$26\%$.
The corresponding residual correlation is $C_{\rm res}=0.997$--$0.999$ (Supplementary Material).
The model thus captures the global $\cos\theta$ pattern and also reproduces the larger residual deviations at edge and vertex sites [Fig.~\ref{fig:heterogeneity}(c,d)].
The morphology-resolved surface charge arises from placing the calibrated response on the discrete atomic sites rather than on a smooth boundary.
This is the atomistic information absent from a continuum boundary-element description at the same particle size.\cite{bem_plasmonics_review} A site-dependent static electronic response on noble-metal nanoparticle facets has also been reported in many-body (\textit{GW}) calculations of charge-injection energies,\cite{lei2023chargeinjection} for particles far smaller than reached here.

\subsection{Computational scaling to experimental sizes}
\label{sec:scaling}

The purpose of an atomistic response model is to reach particle sizes that are inaccessible to first-principles methods, so its cost as a function of system size is as important as its accuracy.
The linear response is obtained by solving the $4N$-dimensional linear system (Methods) with a conjugate-gradient (CG) iteration, the cost of which is dominated by one matrix--vector product with the hardness matrix $G$ per step.
The long-range Coulomb part of this product is evaluated by the continuous fast multipole method (CFMM).\cite{Lazarski2015}
The KXC overlap term is short-ranged and contributes only to the near field, so it adds no asymptotic cost and leaves the per-iteration work essentially unchanged from the uncorrected model.
We therefore take the wall time per CG iteration as the measure of the matrix--vector cost.
Its absolute value depends on the hardware, but its growth with $N$ is the scaling quantity of interest.

\begin{figure}[t]
\centering
\includegraphics[width=\columnwidth]{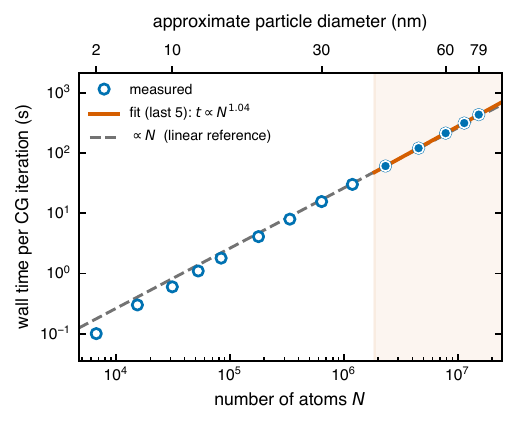}
\caption{Computational scaling of the KXC linear solve. Wall time per conjugate-gradient iteration (one CFMM-accelerated matrix--vector product) as a function of the number of atoms $N$; the top axis gives the corresponding particle diameter (in nm). Circles are measured timings. Filled points in the shaded region ($N\gtrsim2\times10^{6}$) are the asymptotic subset used for the power-law fit (solid line, $t\propto N^{1.04}$). The dashed line is an ideal linear ($\propto N$) reference. Open points below the shaded region, where fixed overhead contributes to the per-iteration cost, are excluded from the fit. The measured cost is near-linear across the full range, reaching a $1.5\times10^{7}$-atom ($\approx$79 nm) particle, far beyond the size accessible to first-principles methods.}
\label{fig:scaling}
\end{figure}

Figure~\ref{fig:scaling} shows this cost for model spherical particles spanning $N=6.7\times10^{3}$ to $1.5\times10^{7}$ atoms (diameters of 6 to 79~nm).
In the large-$N$ regime the per-iteration time scales as $N^{1.04}$, close to the ideal linear cost of a fast-multipole evaluation and far below the $N^{2}$ complexity of direct pairwise summation.
The small residual departure from unit slope reflects the near-field octree work at finite atom counts.
All timings were measured on a single AMD Ryzen Threadripper PRO 5995WX workstation (64 cores) using 64-core OpenMP parallelization.
These sizes lie three to four orders of magnitude beyond the few-hundred-atom ceiling of routine Kohn--Sham DFT.
Comparable atom counts have recently been reached in frequency-domain atomistic plasmonics with preconditioned iterative solvers.\cite{lafiosca2021million}
The present approach complements those developments by targeting the static response and by calibrating and validating the induced density itself against DFT.
Because the KXC correction changes the entries of $G$ but neither its dimension nor its asymptotically dominant long-range structure, the corrected model inherits this scaling at essentially the same per-iteration cost as the uncorrected one.
Three size scales should be kept distinct.
Accuracy is validated against DFT only at the cluster scale, against the 55--101-atom references.
The conductor-crossover diagnostics (Fig.~\ref{fig:large_particle}) are reported as model predictions up to 18~nm.
The matrix--vector cost is shown here to remain near-linear out to 79~nm ($1.5\times10^{7}$ atoms).
With accuracy anchored at the cluster scale, the same fixed parameterization is computationally applicable across the 2--40~nm-diameter window relevant to single-particle plasmonic experiments\cite{sonnichsen2002damping,olson2015singleparticle,campos2019qse} and beyond.
The larger-size results are predictions of the calibrated model.

\section{Discussion}
\label{sec:limitations}

We have developed a static, density-calibrated atomistic description of metallic-nanoparticle screening that connects the DFT-accessible cluster regime to conductor-like integrated behaviour at experimentally relevant sizes.

A model calibrated only to the dipole moment or the polarizability leaves the spatial structure of the surface charge undetermined, because many profiles share the same integral.
Fitting the induced-density profile, together with the short-range kinetic-exchange-correlation hardness that a purely electrostatic kernel omits, recovers the finite width and height of the surface screening layer.
The resulting element-averaged parameters generalize to clusters withheld from their construction and transfer to the tested bimetallic Ag/Au particles, where the core--shell cases mainly probe the Ag shell and the alloy and nanorod test exposed Au surface sites.
The external induced potential and the cluster polarizabilities are candidate observables for experimental tests by off-axis electron holography at demonstrated single-charge sensitivity\cite{gatel2013counting,aso2022science} and molecular-beam electric deflection.\cite{heiles2014deflection,knight1985polarizability}
This density-first, physically parameterized route is complementary to data-driven models that learn the induced electron density from first-principles training data\cite{lewis2023densityresponse,rossi2025vfield} and to frequency-domain atomistic-electrodynamics models parameterized from bulk or optical data;\cite{giovannini2022wfqfmu,nicoli2023bimetallic} its distinguishing feature is calibration to the static DFT induced-density profile with four parameters per element.

Across particle size, with the same parameters, the integrated response approaches the macroscopic conductor\cite{jackson_electrodynamics}: the interior field is nearly cancelled, the angularly binned surface charge approaches $\sigma\propto\cos\theta$, and $\alpha/R^3\to1$.
The response nevertheless never becomes a structureless continuum.
The screening layer keeps a finite width of about one atomic layer, and the surface charge remains morphology-resolved, with vertices departing several-fold more from the conductor law than facets, and edges in between.
This atomic-scale structure, absent by construction from smooth-boundary continuum models, is retained across the crossover sizes examined here.

The model's scope has clear limits.
Accuracy is established relative to the DFT references used for calibration (PBE with def2-TZVP bases), not to experiment, and at the largest reference sizes rather than separately at the nanometre scale; it therefore inherits any systematic error in those references, including the PBE description of noble-metal $d$ bands and exchange--correlation screening.
The element-averaged parameters carry finite statistical uncertainty, reported as cluster-to-cluster scatter and bootstrap intervals in the Supplementary Material, and the predictive test is the leave-one-cluster-out validation above.
The largest residuals occur for faceted and bimetallic targets, consistent with the model's main simplification that each element has one width and one coupling, independent of coordination.
Because the model already reproduces this site heterogeneity from the geometry alone (Fig.~\ref{fig:heterogeneity}), the site dependence is captured without site-specific parameters; a coordination-dependent parameterization\cite{chen2015cddim} could refine the residuals further.
The parameters are accordingly effective static-response parameters for isolated, neutral, compact Ag/Au nanoparticles within the present representation and DFT reference level.
The model also carries no electronegativity or chemical-potential term, so it captures the field-induced response but not ground-state charge transfer between Ag and Au.
Because the finite-field reference removes the same field-free density the comparison remains consistent, and the bimetallic tests show that the field-induced response is reproduced without an explicit charge-transfer term for the systems examined here.
Finally, the model treats the static linear response, not optical spectra or plasmon dynamics; a frequency-dependent extension can build on a density representation whose surface-charge position, width, and external potential are validated against the DFT references and which recovers conductor-like integrated behaviour as size grows, the regime in which time-dependent DFT has mapped the molecular-to-plasmonic evolution of the optical response for Ag and Au clusters.\cite{chaudhary2024ag,chaudhary2026au}

\section{Methods}
\label{sec:theory}

\subsection{Charge--dipole model}
\label{sec:theory_density}

We describe the linear static response of an isolated, non-periodic $N$-atom metallic nanoparticle with the charge--dipole electrostatic model of Bodrenko et al.,\cite{Bodrenko2012} the periodic charge--dipole electrostatic model (PCDEM), applied here in its non-periodic form to finite clusters. The model is classical. It has no explicit kinetic-energy or exchange--correlation term, and no electronegativity or chemical-potential term, so charge redistribution is driven only by the external field and opposed by the electrostatic interactions.

Each atom $i$ at position $\mathbf R_i$ carries an induced charge $q_i$ and dipole $\boldsymbol\mu_i$, built from atom-centred Gaussian basis functions $g_{i\alpha}(\mathbf r)$ of element-specific widths $R_{q,i}$ (the charge channel $\alpha=0$) and $R_{p,i}$ (the dipole channels $\alpha=1,2,3$). The induced charge density is
\begin{equation}
\Delta\rho(\mathbf r)=\sum_{i=1}^{N}\left[\rho_{q,i}(\mathbf r)+\rho_{\mu,i}(\mathbf r)\right],
\label{eq:total_density}
\end{equation}
where the charge density is $\rho_{q,i}=q_i\,g_{i0}$ and the dipole density is $\rho_{\mu,i}=\sum_{\alpha=1}^{3}\mu_{i,\alpha}\,g_{i\alpha}$ (explicit Gaussian forms in the Supplementary Material). The total induced dipole moment follows directly,
\begin{equation}
\mathbf p=\int \mathbf r\,\Delta\rho(\mathbf r)\,d^3r=\sum_{i=1}^{N}\left(q_i\mathbf R_i+\boldsymbol\mu_i\right),
\label{eq:dipole_moment_model}
\end{equation}
and is monitored as an independent check on the parameterization, not as a fit target.

The induced charges and dipoles are the equilibrium solution of the model. The total energy is a quadratic functional of the induced variables, comprising an electrostatic self- and mutual-interaction among the basis functions, with symmetric matrix elements $G_{ij}^{\alpha\beta}$ over atoms $i,j$ and channels $\alpha,\beta$, plus a coupling to the external potential. Because all basis functions are Gaussians, every interaction integral, including the on-site self-interaction, is analytic, so no damping or cutoff is needed. Minimizing the energy subject to charge conservation $\sum_i q_i=0$ gives a symmetric, positive-definite linear system of dimension $4N$, which we solve by the conjugate-gradient (CG) method. The energy functional, the interaction matrix $G$, and the linear system are standard and are given in full in the Supplementary Material.

\subsection{Kinetic-exchange-correlation correction}
\label{sec:theory_kxc}

In the uncorrected model, the interaction matrix $G$ contains only the classical Coulomb interaction between the Gaussian distributions. This is the electrostatic approximation to the density functional theory hardness (response) kernel,\cite{parr1983hardness,parryang1989dft,Bodrenko2013}
\begin{equation}
\eta(\mathbf r,\mathbf r';n_0)=\frac{1}{|\mathbf r-\mathbf r'|}+\underbrace{\frac{\delta^2 T_s[n]}{\delta n(\mathbf r)\,\delta n(\mathbf r')}+\frac{\delta^2 E_{xc}[n]}{\delta n(\mathbf r)\,\delta n(\mathbf r')}}_{\displaystyle \eta_{\mathrm{kxc}}(\mathbf r,\mathbf r';n_0)}\Bigg|_{n=n_0},
\label{eq:hardness_kernel}
\end{equation}
whose first term is the long-range Coulomb interaction and whose remaining two terms, the second functional derivatives of the non-interacting kinetic ($T_s$) and exchange--correlation ($E_{xc}$) energies, constitute the short-range kinetic-exchange-correlation (KXC) contribution.\cite{Bodrenko2013}
Following Bodrenko and Della Sala,\cite{Bodrenko2013} we approximate the KXC part of each matrix element by an extended-H\"uckel-type form proportional to the overlap of the two basis functions,
\begin{equation}
G_{ij}^{\alpha\beta}\approx \underbrace{\iint \frac{g_{i\alpha}(\mathbf r)\,g_{j\beta}(\mathbf r')}{|\mathbf r-\mathbf r'|}\,\mathrm d^3r\,\mathrm d^3r'}_{\text{Coulomb}}\;+\;\frac{k_i^{\alpha}+k_j^{\beta}}{2}\,O_{ij}^{\alpha\beta},
\label{eq:G_kxc}
\end{equation}
with the overlap integrals
\begin{equation}
O_{ij}^{\alpha\beta}=\int g_{i\alpha}(\mathbf r)\,g_{j\beta}(\mathbf r)\,\mathrm d^3r,
\label{eq:overlap}
\end{equation}
which are again analytic for Gaussian basis functions (closed forms in the Supplementary Material).
Here $k_i^{0}\equiv k_{q,i}$ and $k_i^{1,2,3}\equiv k_{p,i}$ are element-specific KXC coefficients for the charge and dipole channels, respectively.
In the local infinite-jellium limit, the coefficients reduce to a single constant $k=\eta_{\mathrm{kxc}}(n_0)$ common to both channels.\cite{Bodrenko2013}
For atoms in a real environment $k_{q,i}$ and $k_{p,i}$ absorb the non-uniformity of the reference density and the quantum (beyond-electrostatic) part of the response.
Setting $k_{q,i}=k_{p,i}=0$ recovers the uncorrected charge--dipole model defined above.
The linear system is solved with the corrected $G$.
Because the overlap integrals decay much faster with interatomic distance than the Coulomb interaction, the KXC term is short-ranged and adds no penalty to the asymptotic cost of the matrix--vector products.

The physical content of the correction is most transparent for a single isolated atom, for which Eq.~\eqref{eq:G_kxc} yields a static dipole polarizability\cite{Bodrenko2013}
\begin{equation}
\alpha=\frac{3\sqrt{\pi/2}\,R_p^3}{1+3k_p/(4\pi R_p^2)}.
\label{eq:alpha_kxc}
\end{equation}
Here $R_p$ sets the spatial extent of the induced density and $k_p$ rescales its magnitude: at fixed $R_p$, increasing $k_p$ lowers the polarizability. In the uncorrected model ($k_p=0$) both roles fall on $R_p$ alone, so fitting the polarizability drives $R_p$ to an unphysically small value and distorts the density.\cite{Bodrenko2013} The KXC term separates the two, giving an independent handle on the response magnitude while keeping $R_p$ physical.

\label{sec:compdetails}

\subsection{Numerical solution}
\label{sec:numsol}
\label{sec:theory_numerics}

Each CG iteration requires one matrix--vector product with $G$.
When the Gaussian Coulomb integrals are evaluated by direct summation, this product scales as $\mathcal{O}(N^2)$.
For large $N$ the cost becomes prohibitive, and we replace the direct summation with a continuous fast multipole method (CFMM)\cite{Lazarski2015}, which lowers the asymptotic scaling well below the quadratic cost of direct evaluation (timings in Fig.~\ref{fig:scaling}).
Production solves are iterated to a relative residual norm of $10^{-10}$, with convergence tests confirming that the observables of interest are already stable at looser thresholds (Supplementary Material).

\subsection{DFT reference calculations}
\label{sec:dft_reference}

DFT reference data were generated with Kohn--Sham DFT in TURBOMOLE (development version V8.0),\cite{turbomole2023} using the PBE exchange--correlation functional\cite{pbe_perdew} and def2-TZVP basis sets\cite{basis_sets_ahlrichs} with the associated effective core potentials for Ag (28-electron core) and Au (60-electron core).\cite{andrae1990ecp} The 55-atom Mackay-icosahedral \ce{Ag55} and \ce{Au55} are the primary calibration references. The element-averaged parameters use an additional set of compact Ag and Au clusters of 55--101 atoms in decahedral, truncated-cube, truncated-octahedron, and octahedral morphologies (six Ag and five Au clusters in total, including the two icosahedra). The bimetallic set contains the \ce{Au13}@\ce{Ag42} core--shell cluster in four morphologies and two mixed systems in which both elements reach the surface, a 55-atom \ce{Au27Ag28} alloy and a 60-atom \ce{Ag48Au12} nanorod. Monometallic geometries are taken from the CSIRO nanoparticle data sets\cite{barnard_agnp,barnard_aunp} and relaxed with density-functional tight binding; the bimetallic structures are model geometries relaxed with an effective-medium potential\cite{jacobsen1996emt} (full cluster list and reference quality in the Supplementary Material). The DFT reference and the atomistic model use identical coordinates, centred at the origin, and the model density and potential are evaluated analytically at the grid points with no interpolation.

For each cluster, self-consistent Kohn--Sham calculations were performed in a weak uniform external field of amplitude $E_0=10^{-4}~\mathrm{a.u.}$ applied independently along $x$, $y$, and $z$, and the induced charge density was obtained by central finite difference,
\begin{equation}
\Delta\rho^{\mathrm{DFT}}_{\alpha}(\mathbf r)=\tfrac{1}{2}\!\left[\rho(\mathbf r;+E_0\hat{\boldsymbol\alpha})-\rho(\mathbf r;-E_0\hat{\boldsymbol\alpha})\right],
\label{eq:finite_diff}
\end{equation}
which cancels the field-even terms and isolates the linear response, with the sign convention that the induced dipole is parallel to the applied field. The induced density was sampled on a uniform $100^3$ grid (about $38~a_0$ box edge) enclosing the cluster, from which we extracted the induced dipole moment $\mu_{\alpha}^{\mathrm{DFT}}$ and the plane-integrated density profile $P_{\alpha}^{\mathrm{DFT}}(u)$ along each field direction.

Because the linear response is a small difference of two densities at weak field, it is sensitive to the self-consistency thresholds. We used tight energy and density convergence ($10^{-8}$~a.u.). With looser convergence the icosahedral clusters, whose polarizability tensor is isotropic by symmetry, develop a spurious direction dependence of up to $15\%$ in the induced response; the tight settings reduce this residual anisotropy below $0.5\%$, and a direct field-doubling test confirms the response is linear at this amplitude. Solver settings, CFMM parameters, conjugate-gradient tolerances, and analytic cross-checks of the post-processing chain are collected in the Supplementary Material; the large-particle timings (Fig.~\ref{fig:scaling}) refer to a single field direction.

\subsection{Error metrics and screening diagnostics}
\label{sec:metrics}
Parameters are selected by matching the plane-integrated induced-density profile $P_\alpha(u)=\iint\Delta\rho\,\mathrm{d}v\,\mathrm{d}w$ (field along $\alpha$) to its DFT reference. The fitting objective is the one-dimensional integrated-square-density (1D-ISD) error,
\begin{equation}
\Delta_{\text{1D-ISD}}^{\alpha} = \sqrt{\frac{\sum_n \left(P_{\alpha}^{\mathrm{DFT}}(u_n) - P_{\alpha}^{\mathrm{model}}(u_n)\right)^2}{\sum_n \left(P_{\alpha}^{\mathrm{DFT}}(u_n)\right)^2}},
\label{eq:1disd_error}
\end{equation}
the relative root-mean-square (RMS) deviation of the model profile from DFT, averaged over $\alpha\in\{x,y,z\}$. The relative dipole error,
\begin{equation}
\Delta_{\mathrm{dip}}^{\alpha} = \frac{|\mu_{\alpha}^{\mathrm{model}} - \mu_{\alpha}^{\mathrm{DFT}}|}{|\mu_{\alpha}^{\mathrm{DFT}}|},
\label{eq:dipole_error}
\end{equation}
with $\mu_{\alpha}$ the induced dipole moment, is monitored as an independent global-response check and is not part of the objective.

Three diagnostics characterize the approach to conductor-like screening at large size, all evaluated from the converged model charges and internal field and defined in the Supplementary Material. The interior screening factor $S$ measures how completely the applied field is cancelled inside the particle core, with $S=1$ the perfect-conductor limit; it is a projection-RMS form of the interior-screening measure of Sinha-Roy et al.\cite{sinharoy2020,sinharoy2023} The conductor-law residual $\epsilon_{\cos}$ is the relative deviation of the angle-binned surface charge from the conductor law $A\cos\theta$. The screening-layer width $w_r$ is the charge-weighted radial spread of the induced charge about its mean radius; it includes the atomistic surface corrugation and the finite extent of the atom-centred basis and is not a frequency-dependent Feibelman width. For the ratio $\alpha/R^3$, $R$ is the construction radius of the model fcc sphere, equal to half the diameter reported in Fig.~\ref{fig:large_particle} and in the crossover table of the Supplementary Material.

\subsection{Parameter optimization}
\label{sec:paramopt}

The parameters are selected by minimizing the direction-averaged 1D-ISD error, which targets the spatially resolved density: the two widths $(R_q,R_p)$ for the uncorrected model, and the four parameters $(R_q,R_p,k_q,k_p)$ for the KXC-corrected model, scanned by coarse-to-fine refinement. The dipole error is not part of the objective and is reported as an independent global-response check. For the finite clusters considered here the four-dimensional 1D-ISD landscape has a well-isolated minimum, as a random multi-start search over the parameter box found no lower error than the coarse-to-fine optimum (Supplementary Material). We therefore retain all four parameters as independent, rather than imposing the $R_q=R_p$, $k_q=k_p$ constraints that Bodrenko and Della Sala required for periodic slabs.\cite{Bodrenko2013}

\subsection{Use of artificial-intelligence tools}
Artificial-intelligence (AI) assistance was limited to language editing. A large language model (Claude Opus 4.8, Anthropic) was used for minor refinements of parts of the text to improve readability.
All AI-modified text was thoroughly scrutinized by the authors and verified to be factually correct, and the authors take full responsibility for the content of this work.
No images, data, or analysis were generated with AI tools.

\section*{Data availability}
The data supporting this study are deposited on Zenodo (DOI \href{https://doi.org/10.5281/zenodo.21460849}{10.5281/zenodo.21460849}).
The deposit contains the DFT-derived induced charge-density profiles, the optimized element-averaged and per-cluster parameter tables, the cluster geometries, the result tables underlying every figure and reported number, and the manuscript figures.

\section*{Code availability}
A direct-integration implementation of the PCDEM/KXC solver is included as a stand-alone Python script in the Zenodo repository, and reproduces the production coefficients from geometry and parameters alone.

\section*{Acknowledgements}
The authors thank Nikhil Kumar for technical assistance in carrying out the DFT reference calculations for the alloy and nanorod bimetallic clusters.
Funding: The authors gratefully acknowledge financial support from Deutsche Forschungsgemeinschaft (DFG, German Research Foundation) within the Collaborative Research Centre (NOA - Nonlinear Optics down to Atomic scales, CRC 1375 project no. 398816777, project A4), the Carl Zeiss Foundation within the Breakthrough Program, and TURBOMOLE GmbH.

\section*{Author contributions}
P.J. developed the model implementation, performed the calculations and analysis, prepared the figures and data package, and wrote the manuscript. M.S. supervised the research, secured funding, contributed to the interpretation of the results, and revised the manuscript. Both authors reviewed and approved the final manuscript.

\section*{Competing interests}
M.S. has an equity interest in TURBOMOLE GmbH and serves as its chief executive officer. P.J. declares no competing interests.

\section*{Supplementary Material}
See Supplementary Material for the formal definitions of the error metrics and of the surface-charge-lobe descriptors, the analytic forms of the KXC overlap integrals, the DFT reference-quality assessment (polarizability tensor isotropy under loose and tight convergence, including the additional clusters), the complete per-direction parameterization results for \ce{Ag55} and \ce{Au55}, the multi-start validation of the optimum and the bootstrap uncertainties of the four-parameter fit, the selection of the multi-cluster training set (the molecular small-cluster exclusion and the cuboctahedral-gold $k_q$ degeneracy), the uncorrected-model and per-direction bimetallic transferability data and the insensitivity of the transfer to the gold-core parameters, the full three-dimensional induced-density comparison with central-slice maps, and the per-shell induced-potential errors.

\bibliographystyle{unsrtnat}
\bibliography{references}

\end{document}